# Momentum-resolved electronic band structure and offsets in an epitaxial NbN/GaN superconductor/semiconductor heterojunction

**Tianlun Yu**[1,2,+,*]**, John Wright**[3,+], Guru Khalsa[3], Betül Pamuk[4], Celesta S. Chang[5], Yury Matveyev[6], Thorsten Schmitt[1], Donglai Feng[2,7,8,9], David Muller[5,10], Grace Xing[10,11], Debdeep Jena[10,11,*] & Vladimir N. Strocov[1,*]

1 Swiss Light Source, Paul Scherrer Institut, CH-5232 Villigen PSI, Switzerland

2 Advanced Materials Laboratory, State Key Laboratory of Surface Physics and Department of Physics, Fudan University, Shanghai 200433, China

3 Materials Science and Engineering, Cornell University, Ithaca, NY 14850, USA

4 Platform for the Accelerated Realization, Analysis, and Discovery of Interface Materials (PARADIM), Cornell University, Ithaca, NY, 14853, USA

5 School of Applied and Engineering Physics, Cornell University, Ithaca, NY 14853, USA

6 Photon Science, Deutsches Elektronen-Synchrotron DESY, Notkestr. 85, 22607 Hamburg, Germany

7 Shanghai Research Center for Quantum Sciences,Shanghai 201315, China.

8 Collaborative Innovation Center of Advanced Microstructures, Nanjing 210093, China.

9 Hefei National Laboratory for Physical Science at Microscale,19CAS Center for Excellence in Quantum Information and Quantum Physics, and Department of Physics, University of Science and Technology of China, Hefei 230026, China.

10 Kavli Institute at Cornell for Nanoscale Science, Cornell University, Ithaca NY 14853, USA

11 Electrical and Computer Engineering and Materials Science and Engineering, Cornell University, Ithaca, NY 14853, USA

+: These authors contributed equally.

*: Corresponding authors: T. L. Y.: 15110190028@fudan.edu.cn; D. J.: djena@cornell.edu; V. N. S.: vladimir.strocov@psi.ch.

**Abstract:** The electronic structure of heterointerfaces play a pivotal role in their device functionality. Recently, highly crystalline ultrathin films of superconducting NbN have been integrated by molecular beam epitaxy with the semiconducting GaN. We use soft X-ray angle-resolved photoelectron spectroscopy to directly measure the momentum-resolved electronic band structures for both NbN and GaN constituents of this Schottky heterointerface, and determine their momentum-dependent interfacial band offset as well as the band-bending profile into GaN. We find, in particular, that the Fermi states in NbN are aligned against the band gap in GaN, which excludes any significant electronic cross-talk of the superconducting states in NbN through the interface to GaN. We support the experimental findings with first-principles calculations for bulk NbN and GaN. The Schottky barrier height obtained from photoemission is corroborated by electronic transport and optical measurements. The momentum-resolved understanding of electronic properties elucidated by the combined materials advances and experimental methods in our work opens up new possibilities in systems where interfacial states play a defining role.

**Introduction**

The metal-semiconductor junction - also known as the Schottky junction - is one of the earliest solid-state devices discovered by Braun[1] as early as in 1874. It was used for rectification of electronic current flow even before the discovery of electron[2]. In the 1930s and 40s, Schottky[3], Mott[4], and Bethe[5] uncovered the reason for the characteristic current rectification using the then-newly formulated quantum mechanical picture of electronic bands in metals and semiconductors. The offset in electron energies between the Fermi level ($E_F$) of the metal and the conduction band maximum (CBM) of the semiconductor was found to be responsible for the very large rectification of electronic current flow across the junction. In spite of the long history, Schottky junctions remain at the forefront of many technological applications ranging from high-speed terahertz electronics to low-power digital electronics and high-voltage power electronics[6–9]. Latest generations of these devices are based on newer families of semiconductors and metals or superconductors, often using quantum confinement and correlation effects in low-dimensional structures. Even with the ubiquity of Schottky junctions in modern electronics, a full quantum mechanical understanding of their workings remains far from complete. An important and still elusive piece is the momentum (**k**) dependent electronic band alignment across the junction as characterized by their individual electronic dispersions $E(\mathbf{k})$.

Angle-resolved photoemission spectroscopy (ARPES) provides the most direct pathway for scrutiny of electron states resolved in **k**-space. Its use to probe buried interfaces is however hampered by a relatively small probing depth of typically less than 0.5 nm, limited by the photoelectron mean free path[10]. This constraint can be removed by the use of soft-X-ray photon energies ($h\nu$~1 keV), whereby the probing depth increases to several nanometers[11] to access buried interfaces. Such applications of soft-X-ray ARPES (SX-ARPES) have recently been demonstrated, for example, for semiconductor/semiconductor interfaces AlN/GaN in high-electron-mobility transistor (HEMTs) heterostructures[12], ferromagnetic/semiconductor interfaces EuO/Si[13] and EuS/InAs[14] where the band offset is crucial for the spin-injection functionality, ferromagnetic/topological interfaces EuS/Bi$_2$Te$_3$ (ref. [15]), metal/strong spin-orbit coupling semiconductor interfaces Al/InAs and Al/InSb prototypic of the Majorana fermionic systems for quantum computing[16], etc (for a recent review see Ref. [17]). Measurement of **k**-resolved energy states at heterojunctions using SX-ARPES requires interfaces of high elemental and structural purity, and ultra-thin overlayer films to allow for photoelectrons generated at the interface to escape.

Nitride materials host a rich palette of semiconducting and superconducting properties, and also host piezoelectricity, ferroelectricity, and magnetism[18]. Semiconducting GaN and its heterostructures are used in blue light-emitting diodes and high-power electronic devices such as diodes and transistors. Metallic NbN becomes superconducting below 17K, enabling applications such as single-photon detectors[19], bolometers[20], and Josephson junctions[21] as building blocks for low-temperature electronics and quantum computing[22,23]. Although the first study of NbN dates back nearly 100 years[24], its electronic band-structure, surprisingly, has never been measured directly. Recently, NbN was epitaxially integrated with gallium and aluminum nitride[25,26], positioning it as an important material for future electronics integrating superconductivity with semiconducting behavior. Properties of the NbN/GaN heterostructure are attractive for potential applications in new devices for quantum computation. For example, the typically incompatible integer quantum Hall effect and superconductivity was recently found to exist concurrently in this material family[27]. Other properties include reduction of phonon escape time[28] and high operating frequency[29] in hot electron bolometers. Critical to progress in this field is a detailed understanding of the **k**-resolved electronic structure of the interface between NbN and the Group III-Nitride materials.

Here, we explore the **k**-resolved electronic structure of a novel Schottky junction of epitaxially grown ultrathin, nearly lattice-matched superconducting NbN interfaced to the semiconducting GaN. By taking advantage of the larger probing depth of SX-ARPES, we directly measure the **k**-resolved band structure of the highly crystalline NbN overlayer and the band structure on the GaN side of the ultrashallow heterojunction buried ~2 nm below. This information allows us to determine, for the first time, not only the band bending in GaN, but also the Schottky-barrier height and **k**-resolved interfacial alignment of the NbN and GaN band structures, which defines the main quantum-mechanical properties of this metal-semiconductor junction. The experimental band structures and alignments are compared with first-principles calculations. The success of our experimental methodology can be carried over to other epitaxial junctions of metals and superconductors with semiconductors and other heterojunctions.

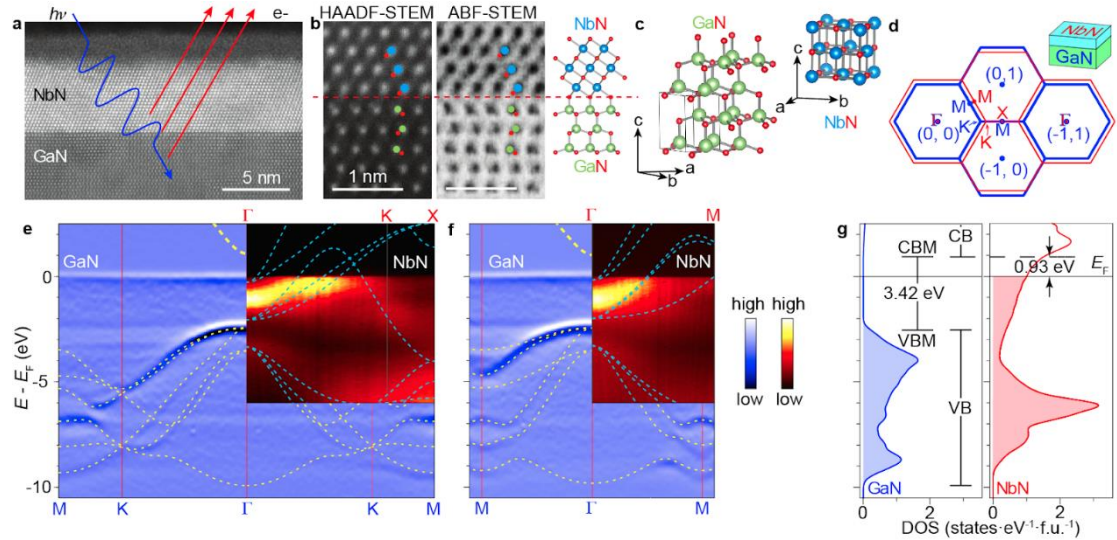

Fig. 1. Characterization of the MBE-grown NbN/GaN heterojunction. **a,** STEM image showing high quality NbN film and sharp interface of NbN/ GaN. A thin oxidized layer (~2 nm) is observed on top of NbN. **b,** Simultaneously acquired HAADF-STEM and ABF-STEM images showing the interface quality and the polarity. Nb, Ga, and N from the atomic ball-and-stick model correspond to blue, green and red spheres, respectively. **c,** The bulk crystal lattice of NbN and GaN. The NbN [111]-direction is aligned with the GaN [0001]-direction in panels **a** and **b**. **d,** Surface BZs of (0001) GaN and (111) cubic NbN. **e-f,** The measured band structure of the NbN/GaN heterojunction along the Γ-K and Γ-M directions. The SX-ARPES intensity from GaN is represented as its second derivative $\partial^2 I/\partial E^2$ shown in blue-white false color, and from NbN as intensity in red-black. The yellow dashed lines are the calculated GaN bands, and the blue ones the calculated NbN bands. **g,** Density of electronic states in bulk GaN and NbN calculated using DFT indicating a barrier height of 0.93 eV.

## Results

**Structural and transport properties of the NbN/GaN heterostructures:**

The structural properties of the MBE-grown NbN/GaN heterojunction with a NbN thickness of 5.5 nm are shown in Fig. 1**a-c**. A high-resolution scanning transmission electron microscopy (STEM) image of the NbN/GaN interface in Fig. 1**a** is superimposed with a schematic of SX-ARPES, illustrating how this technique probes the interface electronic structure for NbN/GaN. Fig. 1**b** shows high-angle annular dark field (HAADF-) and annular bright field (ABF-) STEM image of the GaN/NbN interface in atomic scale. The GaN is metal (Ga-) polar and aligned along the [0001] direction. The NbN [111] axis aligns with the [0001] crystal direction of GaN. Both the Nb and N sublattices can be resolved in Fig. 1**b**. The atomic arrangement of the two crystalline layers is indicated by the ball-and-stick models in Fig. 1**b-c**. The NbN/GaN interface is atomically sharp, with no evidence of disorder, contamination, or intermixing across the interface.

Introducing our electronic structure results, Fig. 1**d** sketches the **k**-space configuration as a superposition of the surface Brillouin zones (BZs) of (0001) GaN and (111) cubic NbN. Fig.

1**e-f** shows the essentials of our SX-ARPES results on the **k**-resolved band structure of the NbN/GaN heterointerface, represented as the experimental band structure of GaN matched to that of the NbN overlayer. In the GaN band structure, the valence-band maximum (VBM) is at the Γ-point, and the dispersions along both the Γ-K-M and the Γ-M directions are clearly resolved. The dashed lines are the energy bands calculated using density-functional theory (DFT), where the unoccupied conduction bands of GaN, inaccessible to SX-ARPES, are shown as a thicker dashed line. For NbN, the experimental band structure shows bands dispersing upwards from the Γ point and crossing the Fermi level ($E_F$) at about halfway along the Γ-K and Γ-M lines. The theoretical bands are superimposed as the dashed lines. One of the key results of our study is that the valence-band offset (the energy difference between the VBM of GaN and $E_F$ of NbN) is determined to be 2.49 eV. With the fundamental GaN band gap of 3.42 eV, this figure implies a conduction-band offset of 0.93 eV at the Γ point of the interface. This is summarized in the density-of-states calculations in Fig. 1**g**. We emphasize that the VBM of GaN is separated in **k**-space from the superconducting states of NbN at $E_F$ by about a quarter of the BZ. We will expand on the SX-ARPES measurements and its implications for the physics of the NbN/GaN heterojunction later on, after describing our measurements of the structural, transport and optical properties of our NbN/GaN samples.

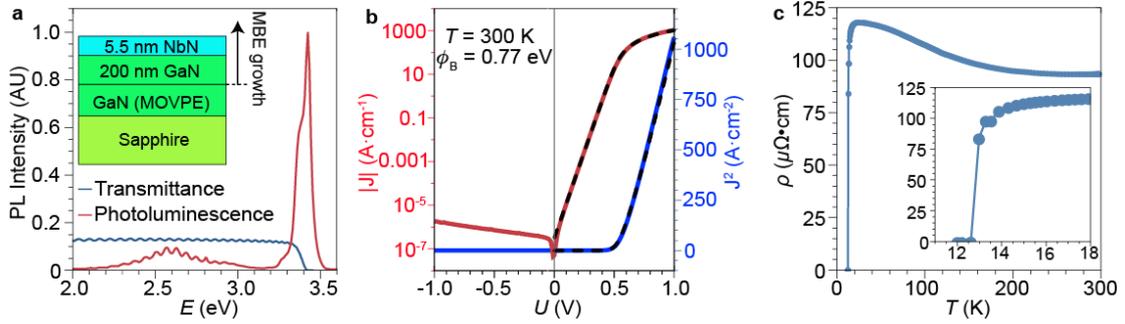

Fig. 2: Properties of the NbN/GaN films. **a,** Photoluminescence and optical transmittance measurements of a 5.5 nm NbN on GaN sample performed at 300K. **b,** IV data for a NbN/GaN Schottky barrier diode performed at 300 K. The dashed line represents the best-fit thermionic emission model with a series resistance. **c,** Resistance versus temperature of a 2.8 nm NbN on GaN film. A clear and sharp transition to the zero resistance state is seen around 13 K.

A series of NbN thin films of differing thickness were grown on GaN to ensure that the NbN would retain its structural and electronic properties when the films were scaled to the thicknesses tailored to the probing depth of the SX-ARPES measurements. The sample series covered the NbN film thicknesses of 1.2, 1.5, 2.0, 2.5, and 10 nm. Reflection high energy electron diffraction (RHEED), X-ray diffraction (XRD), and atomic force microscopy (AFM) were also used to confirm that the NbN films grown on GaN at the thickness necessary for SX-ARPES measurements are epitaxial, with uniform thickness and smooth surfaces. Using XRD we measure the lattice constant of the rocksalt cubic NbN to be $a = b = c \sim 4.34$ Å, while that of the hexagonal GaN is $a = b \sim 3.19$ Å and $c \sim 5.19$ Å (crystal structures are shown in Fig. 1**c**). As a result, the in-plane Nb-Nb spacing is ~ 3.07 Å along (111) orientation, which results in

~3.8 % lattice misfit at the interface. Photoluminescence measurements (Fig. 2**a**) of a sample with a 5.5 nm NbN film on GaN indicates that the MBE grown GaN underneath the NbN exhibits a photoluminescence peak at 3.42 eV. This value is in good agreement with previously reported values for GaN, and provides confidence that the electronic properties of the MBE grown GaN film are preserved through the growth of the NbN thin film, a process which occurs at high temperature. Additional details of the heterostructure growth and characterization as well as removal of the in-situ indium cap needed to protect sample quality until SX-ARPES measurement is described in the Methods section.

Electronic transport across the NbN/GaN interfaces was studied by fabricating circular Schottky barrier diode devices with a diameter of 50 μm. A current-voltage (IV) measurement of such a device performed at 300 K is shown in Fig. 2**c**. The diodes exhibit strong rectification, demonstrating exponential increase in the current of 7 orders of magnitude in forward bias. This behavior is modelled well using thermionic-emission theory,

$$I = aA^{**}T^2 e^{-\frac{q}{kT}\phi_b}(e^{-\frac{q}{\eta kT}(V-IR)} - 1)$$

where $I$ is the current, $a$ is the device area, $A^{**}$ is the Richardson constant for GaN, $T$ is the temperature, $q$ is the elementary charge, $k$ is the Boltzmann constant, $\Phi_b$ is effective barrier height, $\eta$ is the ideality factor, $V$ is the voltage, and $R$ is a resistance in series with the diode. In this model, $\Phi_b$, $\eta$, and $R$ are used as fitting parameters.

Using the GaN effective mass ($m^*$) in the CBM of 0.222 $m_0$ ($m_0$ is the free-electron mass)[30] to calculate the Richardson constant for GaN of 26.64 A $K^{-2}$ $cm^{-2}$, the effective NbN/GaN barrier height is calculated from the best-fit thermionic emission model to be 0.7660±0.0002 eV. Using capacitance-voltage measurements of the same NbN/GaN diode in reverse bias, we determine the donor concentration in the GaN is determined to be ~2*$10^{17}$ $cm^{-3}$. Using this value for the donor concentration in the GaN, the Schottky barrier height lowering due to electric fields within the GaN is calculated[31]. Thereby we determine the fundamental Schottky barrier height of the NbN/GaN junction to be 0.812±0.003 eV. We see a 0.12 eV difference in the measured barrier height when comparing the transport with the SX-ARPES (0.93 eV barrier height) which may be traced to several of the simplifying assumptions of the thermionic emission model, such as the assumption of a spatially homogeneous Schottky barrier, which ignores the effects of threading dislocations in the semiconductor, and the assumption of an absence of tunneling current. Effects confined to within a few monolayers of the interface, such as strain in the NbN and GaN, and chemical bonding between the GaN and NbN, may also affect the electronic transport across the interface without being observable in the SX-ARPES measurement of electronic states. and the. which may trace back to the neglect of band structure effects in the thermionic-emission theory.

Resistance vs. temperature measurements of a NbN film with a thickness of 2.9 nm exhibit both a low normal state resistivity of 94 μΩ cm and a superconducting critical temperature of 12.8 K, as shown in Fig. 2**c**. These values compare favorably to other reports of superconducting and normal metal properties of ultra-thin NbN films, which we ascribe to the high quality of the MBE grown interface and films[32].

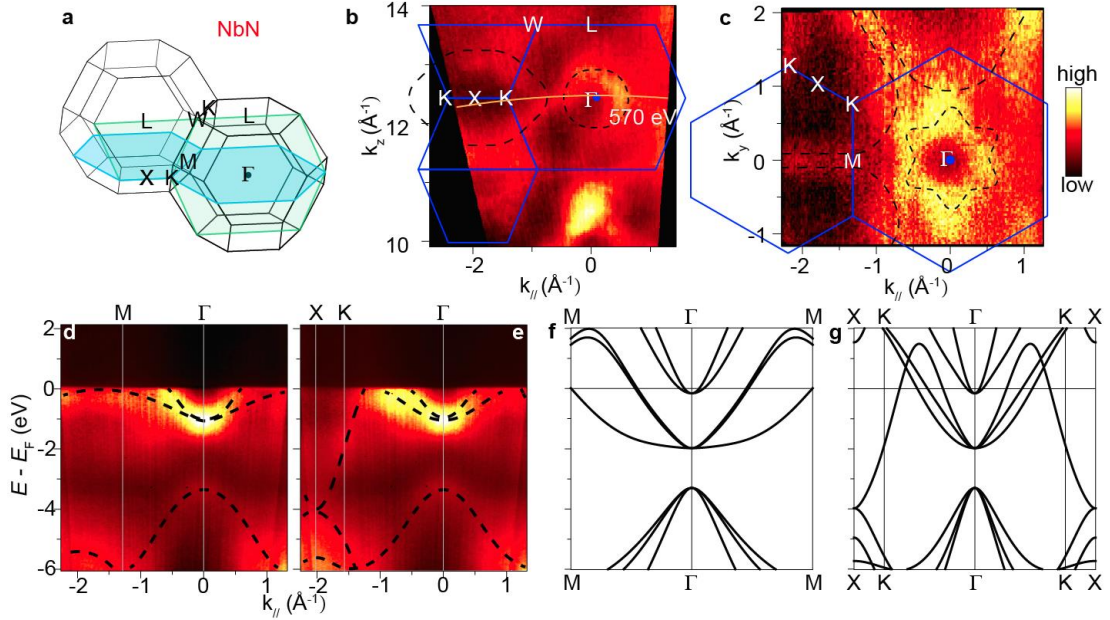

Fig. 3: The electronic structure of NbN. **a**, The BZ of NbN. The green and blue planes indicate the position of the out-of-plane surface (FS) and in-plane FS, respectively. **b-c**, The experimental out-of-plane and in-plane FSs. The orange curve marked by 570 eV indicates the experimental $k_z$ across the Γ point. The dashed lines indicate FS pockets. The blue boundaries represent the BZ edges shown as planes in panel **a**. **d-e**, SX-ARPES intensity along M-Γ and X-K-Γ measured at $hv = 570$ eV. The dashed lines provide a guide to the eye for the band structure and FS cuts in all figures. **f-g**, DFT calculated band structure along the directions measured in panel **d-e**.

**Electronic structure of the NbN/GaN interface:**

Comparing the different samples of the NbN/GaN heterostructure, we find that neither the NbN band structure nor that of GaN shows any changes with the film thickness within the experimental resolution. The highest quality SX-ARPES data on NbN was found for the sample with its largest thickness of 10 nm (Figures 1**e-f** & 3). The GaN band structure could be seen in the $hv$ range of our experiment through a NbN film thickness of less than 2 nm, with the strongest signal recorded in the 1.2 nm thick NbN film (Figures 1**e-f** & 4).

Our results for NbN, the first direct measurement of the **k**-resolved electronic structure of this material, are presented in Fig. 3. The in-plane and out-of-plane Fermi surface (FS) maps (Figs. 3**b-c**, respectively) show large electron pockets of the NbN conduction bands centered at the Γ- and X-points of the extended BZ. The Γ-centered FS pocket in the out-of-plane map has a six-fold symmetry, consistent with the NbN growth direction of [111]. Importantly, there is no spectral weight in the FS of NbN at the Γ-point where the CBM and VBM of GaN are located. The dispersive FS contours in the out-of-plane map identify the three-dimensional (3D) character of the electron states formed in the 10-nm thick NbN film. The broad spectral width of the experimental bands contributing to the FS (Fig. 3**f**-**g**) does not directly allow us to distinguish the number of separate bands, but the band structure calculated from DFT for bulk NbN gives a useful starting point for the interpretation of the SX-ARPES dispersions. An electron-like band centered about Γ with a binding energy ($E_B$) of -1.1 eV is expected to be Nb-3$d$ $t_{2g}$ in orbital character. The octahedral crystal field due to the N-atoms is expected to split the Nb-3$d$ orbitals, pushing the $e_g$ bands up towards $E_F$. The bands starting near $E_B$ = -3.4 eV at the Γ-point are expected to be N-2$p$ in orbital character. DFT underestimated the energy difference between the N-2$p$ and Nb-3$d$ $t_{2g}$ states by ~1 eV, where it is clear that the measured Nb-3$d$ $t_{2g}$ bands are pushed up relative to the N-2$p$ bands. This disagreement could be due to the approximations made in DFT or due to nitrogen vacancies in the NbN film which are known to reduce $E_F$ [33]. The two dispersion branches along Γ-M are consistent between the experiment and theory. The three branches predicted by DFT along Γ-K are not resolved independently in the experimental data, however, the Fermi momentum (halfway the Γ-K line) is consistent between the experiment and theory. The experimental dispersion range of these bands is significantly smaller than that predicted by DFT, which may indicate a strong renormalization of the Nb-3$d$ $t_{2g}$ bands due to yet unknown many-body effects. The increase in the density of states near $E_F$ associated with the enhanced $m^*$ may contribute to the robust superconductivity and high critical temperature known for NbN. While a dispersive band of Nb-3$d$ character can be seen clearly at the K-point, low intensity at the X-point prevents a direct comparison with theory (Fig. 3**e**). DFT predicts that this band changes character from Nb-3$d$ to N-2$p$, becoming degenerate with the N-2$p$ manifold at the X-point (Fig. 3**g**). Overall, our experiment confirms the characteristic features of the DFT band structure for bulk NbN.

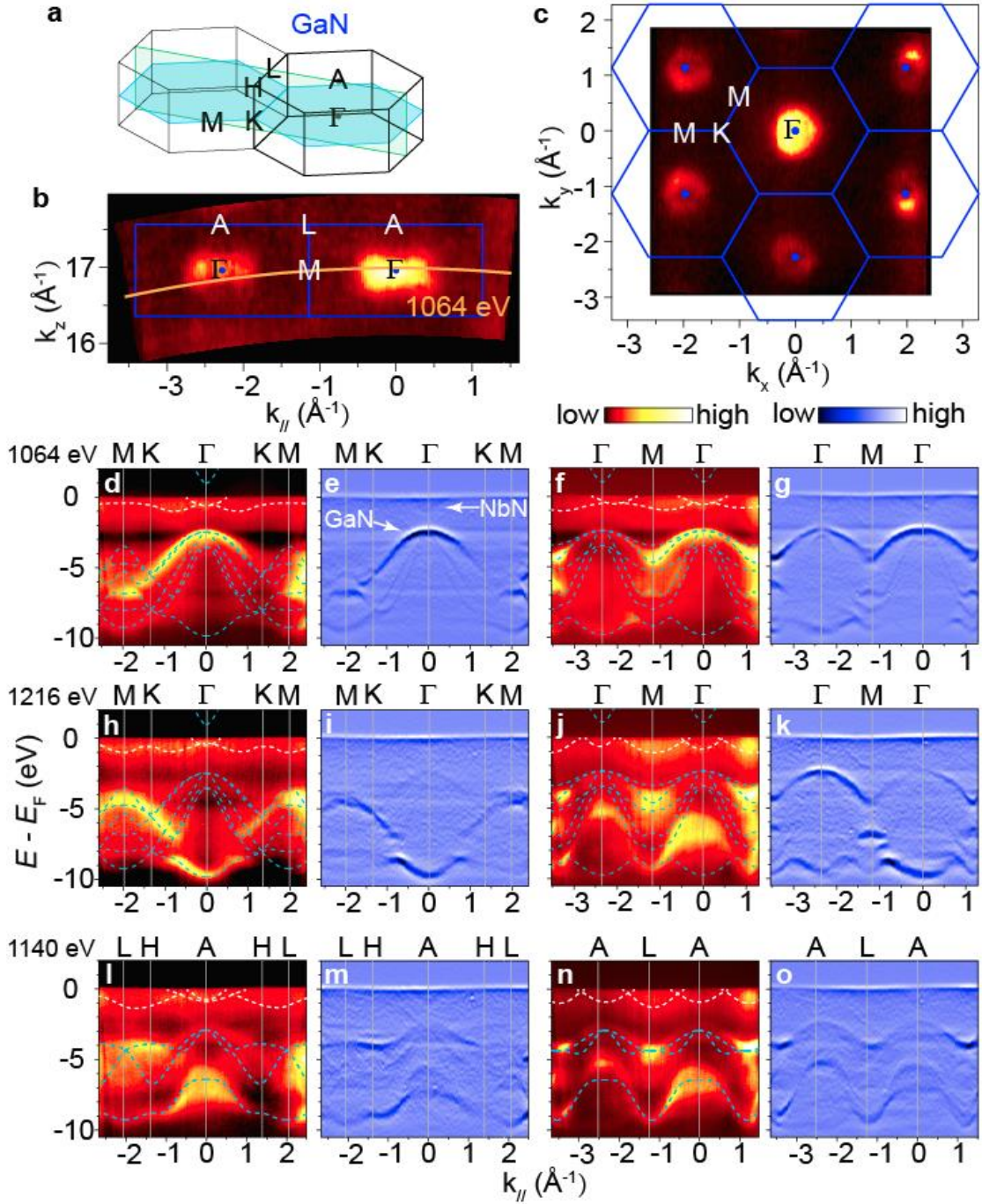

Fig. 4: The electronic structure of GaN. **a,** The BZ of GaN. The green and blue planes indicate the position of the out-of-plane FS and in-plane FS, respectively. **b-c,** The experimental out-of-plane and in-plane iso-energy maps centered at $E_F$ - 2.6 eV. The orange curve marked at 1064 eV indicates the experimental $k_z$ across the Γ point of GaN. **d-o,** SX-ARPES intensity spectrum (red and black) showing the valence bands of GaN along K-Γ-K, Γ-M-Γ, H-A-H and A-L-A at different energies and their corresponding second derivatives (blue and white). The dashed blue lines represent the calculated bands of GaN, and the dashed white lines near $E_F$ trace the experimental NbN bands deviating from those in Figure 3 taken at different $hv$.

We now focus on the GaN band structure measured by SX-ARPES in the 1.2 nm sample (Fig. 4). We note that zero $E_B$ of the heterostructure is defined by $E_F$ of the NbN film. Both in-plane and out-of-plane iso-energy maps taken 2.6 eV below $E_F$ (Figs. 4**b-c**) display the expected hole-like pockets of the Γ-centered valence bands of GaN. The VBM of GaN is 2.49 eV below $E_F$ at the Γ point probed at $hv$ = 1064 eV (Figs. 4**d-k**). Interestingly, although the M-Γ-M and M-K-Γ images taken at $hv$ = 1064 eV (Figs. 4**d-g**) and $hv$ = 1216 eV (Figs. 4**h-j**) correspond to $k_z$ values different by the reciprocal lattice vector 2π/c (where c is the [0001] lattice constant of GaN) and therefore equivalent, the photoemission dipole selection rules in the non-symmorphic crystal structure of GaN for these $k_z$ are different and light up different sets of bands[34]. Apart from an energy shift, the measured band structure is identical to that observed in other GaN-based heterostructures[12]. The calculated bulk band structure of GaN is shown overlaid on the experimental dispersion curves in Fig. 4. The heavy-hole $m^*$ is estimated from a fit to the SX-ARPES data to be ~1.82 ± 0.04$m_0$ compared to the DFT value 2.1$m_0$ (the light-hole $m^*$ can not be evaluated accurately from our SX-ARPES data because this band is much obscured by the heavy-hole band's spectral weight at the Γ-point and, in addition, its apparent dispersion can be flattened by the intrinsic $k_z$ broadening of the ARPES final states[35]). Overall, the agreement between the theoretical and experimental band structure of GaN can be regarded as excellent. Concurrently with the GaN valence bands, our SX-ARPES data resolves NbN conduction bands in vicinity of $E_F$ (dashed white lines in Fig. 4**d-o**). Their difference from the NbN bands in Fig. 3 indicates the 3D character of the electron states formed in NbN. Indeed, the $hv$ values used in Fig. 4 set $k_z$ to the Γ-points of GaN but miss the Γ-points of NbN because of the different $c$ lattice constants of these materials in our heterostructure. Because of large electron concentration and thus small Thomas-Fermi screening length in NbN, only 4.5 monolayers stacked within the 1.2-nm thickness of the NbN film are already sufficient to form its 3D band structure, with quantum-size effects suppressed by an inhomogeneity of the film thickness.

Finally, in order to visualize the **k**-dependent band alignment, we bring together the NbN experimental band structure measured at $hv$ = 570 eV (Fig. 3) and the GaN one measured at 1064 eV (Fig. 4), both corresponding to $k_z$ tuned to the Γ-point. The energy scale of these band structures is matched via the $E_F$ position. These plots, displayed in Fig. 1**e-f**, are the key result of our work. We directly observe that the energy separation of the GaN states from $E_F$, where the superconducting states of NbN are located, dramatically varies across the BZ, and attains its minimum value of 2.49 eV at the VBM of GaN. Importantly, the superconducting states of NbN are separated in **k**-space from this point as much as about a quarter of the BZ. The consequences of these experimental results for physics of the NbN/GaN heterojunction are discussed below.

**Discussion & Conclusion**

Interfacial band offsets are key ingredients in design of technologically important heterostructures. However, they are not often directly measured[36,37,38] and almost never **k**-resolved. Direct measurement of the band offset between NbN and GaN along with the **k**-resolved band structures of the constituent materials and their heterointerface enables integration of NbN into the general electronic materials framework; this sets the stage for streamlined design, modelling, and understanding of fully integrated NbN-based superconductor/semiconductor devices (Josephson junctions with integrated gain, single-photon detectors, etc). Furthermore, the first direct measurement of the NbN band-structure enables a greater understanding of the properties of this widely studied and utilized material.

The **k**-resolved electronic structure at the NbN/GaN interface, directly determined in our SX-ARPES experiments, bears an important consequence on its functional properties. In its normal metal state, moving electrons from the CBM of GaN in the Γ-point into the superconducting states of NbN would require a larger lateral momentum because NbN does not have electronic states at the BZ center. Superconductivity is an effect dictated by electronic structure within the pair-potential energy (several meV) from $E_F$ of the junction. Since the GaN valence/conduction bands are located hundreds of meV below/above $E_F$, due to the large energy denominator in this second-order-perturbative picture, the hybridization between the superconducting states in NbN at $E_F$ and the GaN states is negligible even in the highest-energy point of the latter at VBM. An equally important factor of the negligible hybridization is a large **k**-space separation of these states of about a quarter of the BZ. The absence of such hybridization is crucial to protect the superconductivity on the NbN side of the interface from potential poisoning by an admixture of the states on the GaN side. This property establishes the NbN/GaN interfaces as an important route towards integration of superconductivity into semiconductor technology for electronic devices with novel functionalities.

On the other hand, the Cooper-pair formation in the conventional superconductor NbN can be affected on the electron-phonon interaction (EPI) side through interfacial cross-talk of the phonon modes. It has been reported that such interfacial EPI is crucial to the enhancement of $T_c$ by almost 20 K in the monolayer FeSe/SrTiO$_3$ (ref. [39]), when compared to the optimally doped bare FeSe[40]. Another example is the LaAlO$_3$/SrTiO$_3$ interface, where the deposition of LaAlO$_3$ increases the LO3 phonon energy in SrTiO$_3$ from 100 to 120 meV[41,42]. In NbN/GaN, there are optical phonon modes stretching from from 18 meV to 92 meV[43]. Interfacial EPI in the NbN side of the interface, if significant, could manifest as replica bands shifted from the quasiparticle bands by the phonon energy[40–42] or as electron-dispersion kinks at this energy[44,45]. Our ARPES experiment could not resolve such signatures of the EPI because of the relatively large width of the spectral peaks, insufficient energy resolution, or fundamentally weak interfacial EPI. These considerations call for studies of monolayer-thick NbN/GaN heterostructures by the VUV-ARPES in an attempt to resolve the effects of interfacial EPI.

New important functionalities of the NbN/GaN junction may be envisaged if Cooper pairs can be injected from NbN to GaN. However, the large ~0.9 eV band offset implies that for such an injection a heavily doped GaN tunnel junction is necessary. For enhanced transparency, the growth of thin intervening layers of graded bandgap InGaN that remove the measured energy band offset are needed. This sort of careful interfacial engineering is critical, for example, to seamlessly interface superconducting and integer quantum Hall states in this superconductor/semiconductor heterojunction.

From a methodological perspective, our results highlight the advantages of SX-ARPES as the only spectroscopic technique capable of directly visualizing **k**-resolved electronic structure of buried interfaces and heterostructures. Importantly, our present study uses tunable synchrotron radiation to determine band structure of NbN and GaN at different photon energies as dictated by their different out-of-plane lattice constants. The direct measurement of the **k**-resolved electronic states across the NbN/GaN interface has been enabled by the fusion of SX-ARPES technique with the development of MBE growth techniques capable of producing crystalline, coalesced NbN films on GaN down to 1.2 nm in thickness. Applications of such ultra-thin epitaxial superconducting NbN films for advanced electronic devices such as single-photon detectors[46] are currently being investigated.

In conclusion, we have measured the momentum-resolved band structure of the all-epitaxial NbN/GaN heterointerface – a system allowing integration of superconductivity into nitride semiconductor technology – using SX-ARPES. Advances in crystal growth and resulting crystal quality of the GaN and NbN heterointerface allowed us to measure the GaN band structure at the interface while concurrently resolving, for the first time, the band structure of the nearly century old superconductor NbN. Both GaN and NbN band structures are well-described by DFT studies of the bulk materials. This study of the electronic structure of the heterointerface gives a direct measure of the band-alignment between two constituent materials. A large separation of the superconducting NbN states from the GaN states in energy and **k**-space excludes their hybridization, thereby protecting the superconductivity from potential poisoning by the GaN electrons. The materials advances, methods, and new momentum-resolved understanding of the electronic properties of heterojunctions presented in this work highlight the power of combining MBE and SX-ARPES and further position nitrides as a scalable, industrially capable platform for future hybrid superconductor/semiconductor electronics.

**Methods**

The NbN/GaN heterostructures were grown by nitrogen plasma-enhanced molecular beam epitaxy (PAMBE) on commercially available hydride vapor phase epitaxy (HVPE) GaN on sapphire wafers. Ga (99.99999% purity) was provided by a Knudsen cell and active nitrogen species were provided using an RF plasma source. The MBE grown GaN films were not intentionally doped but are n-type as is the case for undoped PAMBE grown GaN films. GaN films of approximately 200 nm in thickness were grown by PAMBE on the HVPE GaN. The GaN was grown in Ga-rich conditions and upon completing the GaN growth the substrate temperature was increased to a surface temperature of 710°C to thermally desorb excess Ga that had accumulated on the film surface for 5 min before the NbN was nucleated. Nb (99.95% purity) was supplied by an electron beam evaporator; Nb flux is measured using an electron impact energy spectroscopy (EIES) system measuring the optical emission of the Nb. During growth the nitrogen flux exceeds the Nb flux by a factor of approximately 3.5; the NbN film growth rate is 1.8 nm/min. *In-situ* RHEED measurements during and after growth confirmed that the rocksalt cubic NbN grows in the (1 1 1) orientation. For characterization of the

NbN/GaN heterostructure in atomic-scale, cross-sectional TEM specimen was prepared using a Thermo Fisher Helios G4 UX Focused Ion Beam (FIB), with a final milling step of 5 keV to reduce damage. Carbon and platinum layers were sputtered on the sample surface prior to FIB milling to minimize potential damage. The sample was then examined by scanning transmission electron microscopy (STEM), using an aberration-corrected Titan Themis operating at 300 keV.

To prevent contamination, all NbN films measured by ARPES were capped with indium films of approximately 800 nm in thickness grown prior to removing the samples from the growth reactor. The indium films were grown with the sample at room temperature to aid agglomeration of the film; the growth reactor chamber was in a high vacuum state during indium film growth. AFM was used to confirm that the indium capping film was coalesced.

NbN films on GaN similar to those used for the SX-ARPES measurement were grown without indium capping to allow for access to the NbN films. Uncapped NbN samples develop a surface oxide layer when exposed to air[47], and in this respect the uncapped samples differ from the indium capped samples used for SX-ARPES. The thickness and thereby the growth rate of these uncapped NbN films was measured using x-ray reflectivity (XRR). The thickness of films used for SX-ARPES was not measured directly but determined using the growth rate determined using thickness measurements of the uncapped samples.

The indium capping films were removed via annealing at 690~730 °C by e-beam heating method for 1~2 hrs *in-situ*. SX-ARPES experiments were performed in the photon energy ($h\nu$) range 350-1250 eV at the ADRESS beamline of the Swiss Light Source, Paul Scherrer Institute, Switzerland. The photoelectrons were detected with the analyzer PHOIBOS-150 from SPECS GmbH at an angular resolution of ~0.1°. The combined energy resolution was set to 80-250 meV. The isoenergetic maps were integrated within ±50 meV. The SX-ARPES experiments were performed in vacuum $<9*10^{-11}$ mbar. In order to suppress smearing of the coherent spectral component caused by the thermal atomic motion[48], the experiments were performed at a base sample temperature of ~20 K. All the data shown in this study is probed by *p*-polarization photons.

Density-functional theory (DFT) calculations are performed using the VASP[49,50] code with Projector Augmented Wave pseudopotentials and plane wave basis sets. For NbN, we use the generalized gradient approximation (GGA) as implemented in the Perdew-Burke-Ernzerhof (PBE) functional[51] with Methfessel-Paxton smearing for the occupation of the electronic states of 0.1 eV with an electron-momentum grid of 18X18X18 and a kinetic energy cutoff of 700 eV. For GaN, we use the range-separated hybrid functional HSE06[52] with Gaussian smearing of 0.01 eV with an electron-momentum grid of 10×10×6 and a kinetic energy cutoff of 600 eV. For both materials, atomic coordinates and lattice parameters are relaxed using the PBE functional with a $10^{-3}$ eV/Angstrom force convergence condition. For the energy convergence, a threshold on the change in total energy of $10^{-8}$ eV is used for all calculations. An additional scissor-cut was applied to the calculated GaN conduction band in order to match the experimental band gap measured in this study.

The method to extract the band bending profile is as described elsewhere[53,54]. Core level peaks were modelled as a sum of several reference spectra with their own $E_B$ shift and intensity. The effective attenuation length was calculated using TPP-2M formula[11]. The reference spectra was simulated as an ideal asymmetric Voight doublet, where the peaks area ratio and separation were taken from literature; Lorentz broadening was taken from the optimal fitting of all obtained spectra, and the Gauss broadening was calculated from beamline and spectrometer resolution.

**Acknowledgements**

We thank N. B. M. Schröter for advice on the ARPES data processing. GK and BP acknowledge the National Science Foundation (Platform for the Accelerated Realization, Analysis, and Discovery of Interface Materials (PARADIM)) under Cooperative Agreement No. DMR-1539918.

# Supplementary Information for

# Momentum-resolved electronic band structure and offsets in an epitaxial NbN/GaN superconductor/semiconductor heterojunction

Tianlun Yu & John Wright, *et al*.

**GaN band bending**

The Ga 3d core level shows a clear energy shift as a function of $h\nu$ between 350 eV to 1250 eV as shown in Fig. 5**a**, while the Nb 4s core level not only exhibits an approximately constant peak position, but also identical line shape. This indicates that the band bending occurs in the GaN layer, causing a shift of the Ga 3*d* peak of ~80 meV in our $h\nu$ range (Fig. 5**b**). Using our deconvolution method as described in the Methods section, the $U(z)$ profile was extracted based on the simple approximation $U(z) \sim z^2$ near the interface, as summarized in Fig. 4**c**, assuming that the VBM position at ~ 1 nm from the interface is equal to the experimental value -2.49 eV (Fig. 5**c**). Though the confidence region appears large, the determined $U(z)$ clearly shows an upwards band bending consistent with the downward energy shift of the Ga 3*d* peak. The variation of $U(z)$ by ~300 meV over a distance of ~6 nm from the interface is only ~1/3 of that observed in other GaN-based heterojunctions such as the GaN/GaAlN[1] where $U(z)$ at the interface end is pulled down by the polarization charge and its variation with $z$ is sharpened by a high density of the mobile electrons accumulated in the interfacial quantum well. The smoothness of this $U(z)$ explains why the experimental band dispersions in NbN/GaN appear significantly sharper compared to those in GaN/GaAlN measured at essentially the same $h\nu$.

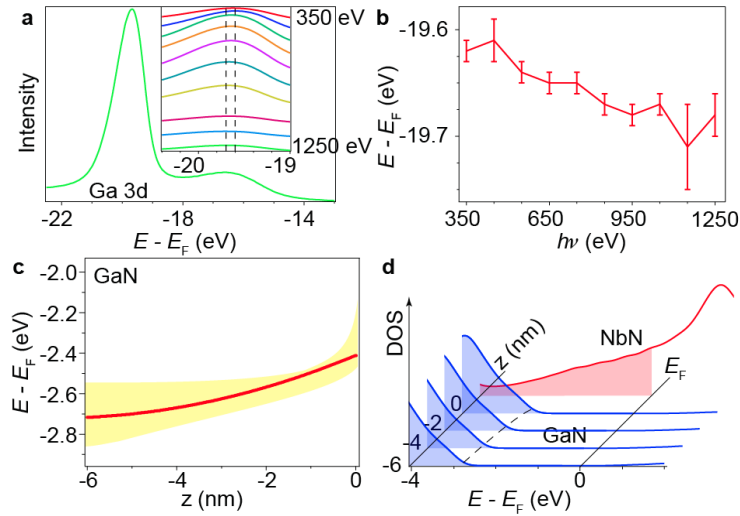

Fig. 5: The band bending in the NbN/GaN heterojunction. **a,** The energy-dependent core level of Ga 3d. **b,** The core level peak position vs $h\nu$. The error bar is determined by the $E_F$ measurements before and after the core level measurement. **c,** The depth profile of the band bending. The yellow shading represents the confidence region and the red line is the optimum profile, assuming that the VBM position at ~ 1 nm from the interface is equal to the experimental value -2.49 eV. **d,** The bend profile of GaN/NbN heterojunction. The CBM of GaN is estimated by the calculation and the band bending profile extracted by experiments.

**GaN band structure comparison**

We present the data from Fig. 3 in the main text with the SX-ARPES and calculated band structure from GaN side-by-side for detailed comparison.

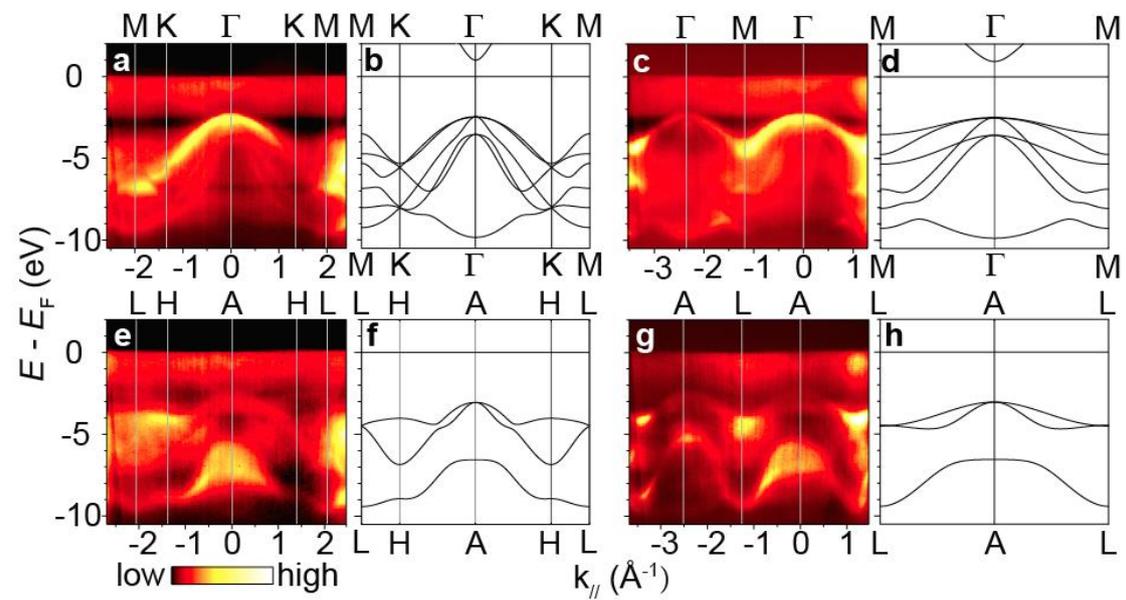

Fig. 6: Comparison between the SX-ARPES and DFT band structure of GaN.

**Reference**

1. Lev, L. L. *et al.* k-space imaging of anisotropic 2D electron gas in GaN/GaAlN high-electron-mobility transistor heterostructures. *Nat. Commun.* **9**, 2653 (2018).